# TITLE

Optimizing the location of vaccination sites to stop a zoonotic epidemic.

# AUTHORS

Ricardo Castillo-Neyra[1,2], Bhaswar Bhattacharya[3], Aris Saxena[3], Brinkley Raynor[1], Elvis Diaz[2], Gian Franco Condori[2], Maria Rieders[3], Michael Z. Levy[1,2]

# INSTITUTIONS

1. Perelman School of Medicine, University of Pennsylvania, Philadelphia, PA, USA

2. Zoonotic Disease Research Lab, School of Public Health and Administration, Universidad Peruana Cayetano Heredia, Lima, Perú

3. The Wharton School, University of Pennsylvania, Philadelphia, PA, USA

# ABSTRACT

**Introduction:** The mainstay of canine rabies control is fixed point mass dog vaccination campaigns. Across Asia, Africa, and Latin America, these annual campaigns consist of dog owners bringing their dogs to stations set up to provide rabies vaccinations. However, in some regions, ideal vaccination coverage in dogs is not obtained due to low participation in the mass dog vaccination campaigns. Travel distance to the vaccination sites has been identified as an important barrier to participation. We aim to increase mass dog vaccination campaign participation by optimally placing fixed point vaccination locations to minimize walking distance of owners to the nearest vaccination location.

**Methods:** We quantified participation probability based on walking distance to the nearest vaccination point using a Poisson regression model. The regression was fit with survey data collected from 2016 to 2019. We then used a computational recursive interchange technique to solve the facility location problem to find a set of optimal placements of fixed point vaccination locations. Finally, we compared predicted participation of optimally placed vaccination sites to historical participation data from surveys collected from 2016-2019.

**Results:** We identified the p-median algorithm to solve the facility location problem as ideal for fixed point vaccination placement. We found a predicted increase in mass dog vaccination campaign participation if vaccination locations are placed optimally. We also found a more even vaccination coverage with optimized vaccination sites; however, the workload in some optimized locations increased significantly.

**Discussion:** We developed a data-driven computational algorithm to combat an ongoing rabies epidemic by optimally using limited resources to maximize vaccination coverage. The main positive effects we expect if this algorithm is to be implemented would be increased overall vaccination coverage and increased spatial evenness of coverage. A potential negative effect could be the presence of long waiting lines as participation increases.

**KEYWORDS:** Access to Health Care, Facility Location Problem, Mass Vaccination, One Health, Optimization, Rabies, Zoonoses.

**INTRODUCTION**

Zoonotic epidemics and pandemics are an increasing public health threat worldwide. In Latin America, Asia, and Africa, epidemics of rabies and other zoonotic diseases are ongoing in major urban centers (1–8). Vaccination efforts to eliminate canine rabies from Latin American countries have been mostly successful (6). However, Peru is experiencing the first instance of canine rabies reintroduction into an area previously declared free of transmission in Latin America (9). In the city of Arequipa and surrounding provinces, continued and increased transmission in the free-roaming dogs (10), the main animal reservoir, has put more than a million human inhabitants at risk of rabies, a fatal, but entirely preventable, disease (11). Annual mass dog vaccination campaigns (MDVCs) have been implemented in Peru to eliminate the epidemic without success (12). The Pan-American Health organization recommends an annual canine mass vaccination coverage of 80% (13); however, reaching this goal has not been attained in the past five years' vaccination campaigns, allowing rabies virus to persist in the free-roaming dog population (12,14,15).

Most MDVCs in Latin America and Africa rely on fixed-location vaccination posts, where vaccinators wait for dog owners to bring their dogs to a set place (14,16–19). The extensive application of fixed-location vaccination is due to its relative ease of implementation and lower cost compared to other strategies (18,20,21). However, in some contexts, fixed-point MDVCs have failed to attain coverage targets (12,22,23). The overarching aim of this study is to explore methods to increase fixed-point MDVC coverage.

Extensive behavioral research has been conducted to reduce refusal of human vaccines (24–29); fewer analogous studies have focused on non-participation in MDVCs (12,14,16,18,19,30,31). Among the barriers reported by dog owners to their participation in MDVCs are inconvenient locations and distance to the vaccination posts (14,16,18). In urban areas, participation in MDVC directly decreases with each city block of distance from the dog owner's household (12). Because rabies virus is transmitted at very low levels within dog populations (32), the virus can persist within pockets of unvaccinated dogs (33–36), making the prospects of elimination more difficult when MDVC do not generate spatial evenness (37).

Despite these reports, in Arequipa, the locations of fixed vaccination posts are mainly determined by convenience and recognizability (e.g. the entrance to a health post, a well-known park) (12). To address this problem, we present an algorithm that provides a data-driven strategy to guide the placement of fixed-point vaccination sites.

The algorithm we present here is derived from the so-called 'facility location problem' and is based on the assumption that the effectiveness of a facility location is determined by some function of the distance traveled by those who visit it (38). With increasing traveling distance, facility accessibility decreases, and thus the location's effectiveness decreases (38,39). This relationship holds for facilities such as libraries and schools, to which proximity is desirable (40), and, based on previous field rabies studies (12,14,18), could also hold for MDVC. Current practice MDVC locations fail to generate spatial evenness in vaccination coverage (12). Using a spatial approach to improve the effectiveness of vaccination sites could not only increase overall vaccination coverage but also create a more even geographic distribution of vaccines. Our objective was to develop an optimization algorithm to determine the placement of vaccination points that would maximize vaccination coverage. We also evaluated vaccination spatial evenness and distribution of workload at vaccination sites. We present a comparison between our algorithm and results from a current-practice MDVC implemented in Arequipa, Peru, in an attempt to quell the dog rabies epidemic in the city.

**METHODS**

**Ethics statement**

Ethical approval was obtained from Universidad Peruana Cayetano Heredia (approval number: 65369), Tulane University (approval number: 14–606720), and University of Pennsylvania (approval number: 823736). All human subjects in this study were adults.

**Data**

Our data were collected from the Alto Selva Alegre (ASA) district in Arequipa city. The Ministry of Health organizes a MDVC to vaccinate dogs against rabies virus every year. Briefly, the Ministry of Health set up and staffed various fixed-location vaccination posts across the district. During planning meetings with the campaign implementers they expressed that vaccination tent locations were selected based on convenience and intuition of the public health officials. A full description of MDVC operations in Arequipa is available elsewhere (12). The data we used for our study consists of two sources: 1) geographic locations of the fixed-location vaccination posts, 2) surveys conducted yearly in ASA immediately following the 2016-2019 MDVC to ascertain house-hold participation in the MDVC.

*1) Vaccination post location:* Fixed-location vaccination posts were georeferenced during the yearly MDVC by our team. All open sport fields, squares, parks, and other open spaces were georeferenced and then approved by MOH health inspectors to ensure they could be used as vaccination sites Out of 85 potential sites, 15 were rejected by Ministry of Health officials as infeasible, leaving 70 potential sites for our optimization analysis (Figure 1).

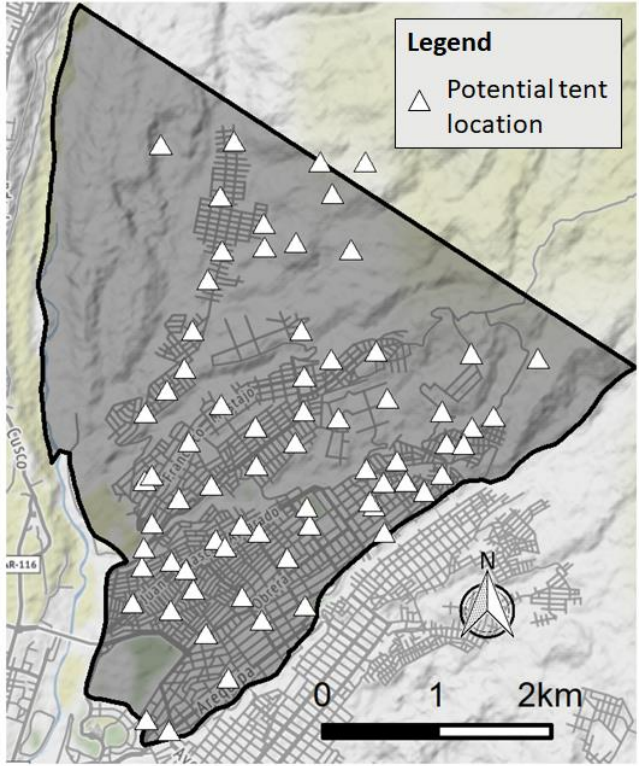

**Figure 1: Alto Selva Alegre district with possible vaccination locations.** These are all potential sites of fixed location vaccination tents. The optimization algorithm selects the optimal locations among these possibilities. The district boundary of ASA is from Peru's National Geo-referenced Data Platform Geo Peru (41).

*2) Vaccination campaign surveys:* a full description of survey methods is available from Castillo-Neyra et al., 2019 (12). Variables collected in these surveys analyzed in this study include: geographic location of each house in the study area, house participation in the MDVC, number of dogs owned by the house-hold, and number of dogs vaccinated in the 2016 MDVC. Surveys following the annual MDVC were conducted from 2016-2019. Data from all four years were used to construct regression models. Data from 2016 were used to demonstrate benefits of optimizing fixed-location vaccination post placement.

**Regression construction of participation probability**

Distance to a fixed-location vaccination point as an important factor influencing a dog owner's participation in a MDVC (12,14). To further explore this relationship, we constructed a model to estimate MDVC participation probability as a function of distance between the house and the MDVC. We calculated the shortest walking distance between each household in the study and the closest fixed-location vaccination point. Shortest walking distances were obtained using Mapbox Directions API and the Leaflet Routing Machine (42,43). Participation of each household in the vaccination campaign was extracted from the 2016 through 2019 MDVC survey data.

We analyzed eight regression models to assess the relationship between participation probability and shortest walking distance from the nearest vaccination point: poisson, negative binomial, binomial using linear distance terms and a combination of linear and quadratic distance terms. For the Poisson and negative binomial regression, we constructed distance bins of 30 meter distances and predicted the number of houses to participate offset by the number of houses per bin. We also constructed mixed-effects versions of each model with a random-effect incorporated per year. All models were evaluated using prediction error (a 70-30 train-test split), AIC, as well as a chi-squared test for the binned data to test if there were significant differences between actual participation (per bin) and predicted participation (Table 1).

**Optimal placement of tents**

We simulated a campaign where vaccination tents were instead placed by solving an extension of the facility location problem: Teitz and Bart's "P-median problem" (44). The aim of the P-median problem is to find a subset of size, *p*, given a set of points, where summed distances of any point in the set to the nearest point in *p* is minimised (44). To apply this to mass dog vaccination in Arequipa, we aimed to find a set of 20 vaccination tent locations, out of a list of all possible tent locations. The number of tents (p=20), was selected to match the number of tents annually run by the MOH in the MDVC. We then applied a greedy heuristic solution of the Teitz and Bart problem (45) in three ways: 1. By minimizing the average travel distance of households to the closest vaccination tent in the study area, 2. By minimizing the maximum distance between any house and the closest vaccination tent, and 3. By maximizing predicted vaccination coverage across the total study area (as predicted by the described probability of participation negative binomial regression). We provided the algorithm with all of the georeferenced house locations in the study area, a matrix of shortest walking distances between all the houses and all of the tent locations and the locations of the potential vaccination points. We assessed optimization placements by examining the distribution of catchment sizes, distributions of walking distance between houses and nearest vaccination point, distribution of estimated probability of vaccination, and estimated overall vaccination coverage.

**Computation and visualization**

All analyses were performed in R (46). We used the MASS (47) and glm packages (48) to fit regressions, the tbart package to optimize tent locations (45), and ggplot2 (49) and ggmap (50) packages to create figures. Base maps for all maps come from OpenStreetMaps (51). Our code used to perform analyses is publicly available on GitHub (52).

## RESULTS

**Regression construction of participation probability**

Our regression analysis included 3463 household surveys from 2015-2019. We constructed and analyzed eight regression models to assess the relationship between travel distance to the closest vaccination tent and probability of participation (Figure 2). Based on lowest AIC, prediction error for out-of-sample dataset and chi square comparison of binned data, we selected the negative binomial regression as the best fit (Table 1).

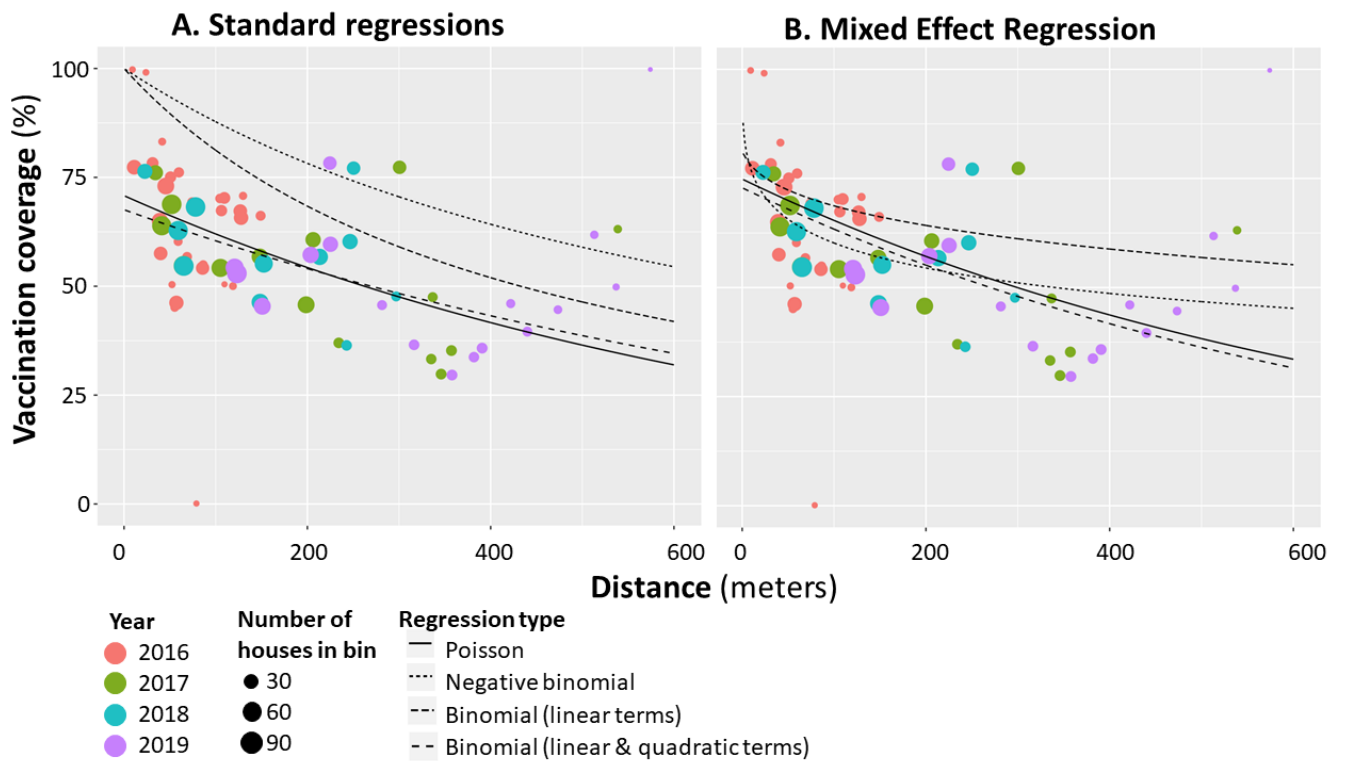

**Figure 2: Regression models of the effect of distance on vaccination coverage.** Regressions constructed comparing the exposure of shortest walking distance to the nearest vaccination tent on the outcome of participation in the vaccination campaign. Houses were grouped in 30 meter bins based on walking distance from the nearest vaccination tent. The percent of houses participating per bin was then compared to the mean distances from all of the houses within a bin to their closest vaccination point.

**Table 1: Regression Analysis Comparison.** All computation was done using the GLM and MASS packages in R. Lambda refers to the number of expected houses participating per 30m bin from the vaccination point, $n_I$ refers to the number of houses per bin, x represents the distance (m) from the nearest vaccination tent (for binned data, mean distance per bin), p represents probability of vaccination (per household), Z (in the mixed models) represents year (2016-2019). The chi square tests for differences between the binned predicted probability (for the binomial models, houses with greater than 50% probability of participation were counted as participating) compared to the survey data of whether or not they participated. Very high p-values provide strong evidence of no significant difference between the predicted and historical distributions.

| Name | GLM Regression equation | Pred. Error | AIC | Chi sq |
|---|---|---|---|---|
| Poisson | $log(\lambda_x) = log(n_i) + \beta_0 + \beta_1 x$ | -0.96% | 123.74 | X-squared = 7.1594, df = 19, p-value = 0.9933 |
| Negative binomial | $log(\lambda_x) = log(n_i) + \beta_0 + \beta_1 x$ | 0.89% | 183.84 | X-squared = 6.9832, df = 19, p-value = 0.9943 |
| Binomial | $logit(p_x) = \beta_0 + \beta_1 x$ | -0.9% | 2837.4 | X-squared = 7.57, df = 19, p-value = 0.9905 |
| Linear + Quadratic terms | $logit(p_x) = \beta_0 + \beta_1 x + \beta_2 x^2$ | -0.96% | 2829.3 | X-squared = 7.57, df = 19, p-value = 0.9905 |
| Negative binomial mixed effects | $log(\lambda_x) = log(n_i) + \beta_0 + \beta_1 x + \mu Z$ | -1.57% | 886 | X-squared = 46.991, df = 159, p-value = 1 |
| Poisson mixed effects | $log(\lambda_x) = log(n_i) + \beta_0 + \beta_1 x + \mu Z$ | -1.57% | 884 | X-squared = 46.991, df = 159, p-value = 1 |
| Binomial | $logit(p_x) = \beta_0 + \beta_1 x + \mu Z$ | -1.00% | 2795.6 | X-squared = 6.3276, df = 19, p-value = 0.997 |
| Linear and Quadratic terms mixed effects | $logit(p_x) = \beta_0 + \beta_1 x + \beta_2 x^2 + \mu Z$ | -1.00% | 2794.8 | X-squared = 6.8283, df = 19, p-value = 0.9951 |

The poisson model was selected for continued analysis because it has the lowest AIC score and equivalent prediction error and chi square p-values. The poisson regression parameters were estimated using the glm package in R (48) and are displayed in Table 2.

**Table 2: Poisson regression parameter estimation**

| Parameter | Estimate | Standard deviation | p-value |
|---|---|---|---|
| 0 | -0.3396406 | 0.0484429 | 2.36e-12 |
| 1 | -0.0013495 | 0.0002549 | 1.19e-07 |

The negative coefficient of $\beta_1$ indicates that as distance increases from a vaccination point increases, expected participation decreases; specifically, for every one meter increase in distance, the expected number of households per distance bin decreases by a factor of 0.999.

**Optimal placement of tents**

Out of the 70 potential tent locations, the 20 most optimal tent locations were selected using three different methods: conventional placement, p-center optimization (minimized maximal walking distance), p-median (minimized overall walking distance) (Figure 3). Catchments are defined as the set of houses around a given vaccination tent for which that tent is the closest via walking distance. The selected tent locations and respective catchment areas for each method are displayed in Figure 3 (Figure 3 A1, B1, C1). From this optimal selection of tent locations, we examined the shortest walking distance distributions (Figure 3 A2, B2, C2). The ideal distance distribution would be shifted towards 0 as much as possible to minimize the distance travelled per household in the study area. The distributions of P-median were shifted towards 0 the most, as one would expect because this method minimizes overall walking distance.

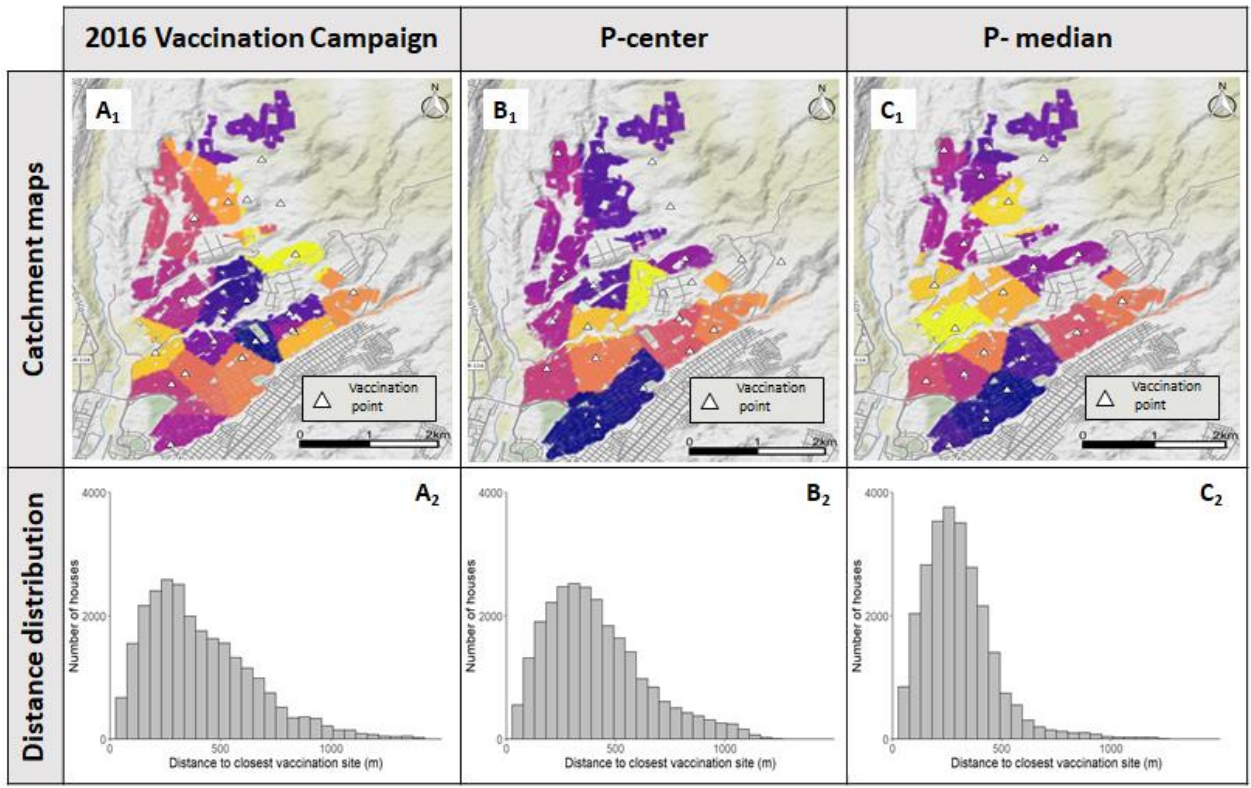

**Figure 3: Vaccination point selection method comparison.** Vaccination tent locations (white triangles) and subsequent catchment areas (different colored regions) are mapped in Panel 1 based on different tent selection methods: convenience (A), P-center: minimized maximal walking distance (B), and p-median: minimized overall walking distance (C). The corresponding histograms in panel 2 depict the distribution of distance to the closest vaccination point per each method.

The p-median method clearly does the best at minimizing overall walking distance. However, we also assessed the workload for each tent location based on the number of houses in the catchment (Figure 4). An ideal distribution of workload would look uniform with the work being spread evenly over the catchements. Again, the p-median algorithm performs the best (Figure 4C).

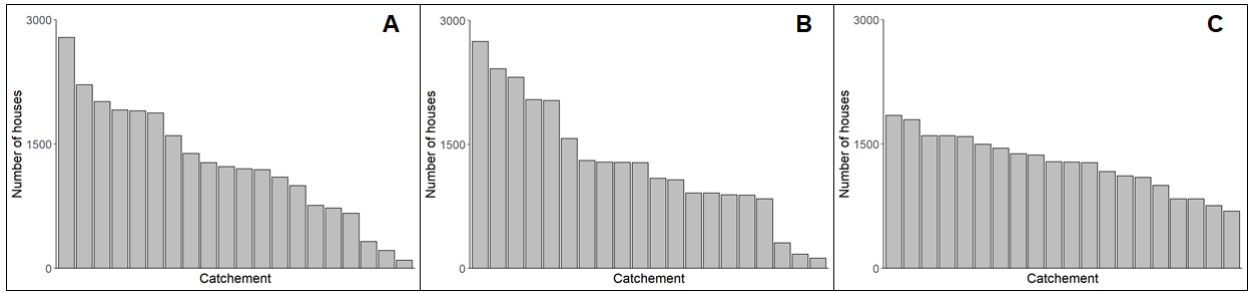

**Figure 4: Distribution of workload across catchment areas.** Number of houses allocated to each catchment are displayed for the 3 different tent placement methods: convenience (A), P-center: minimized maximal walking distance (B), and p-median: minimized overall walking distance (C).

The P-median algorithm did better at both minimizing overall walking distance and evening out workload across vaccination tents. To assess the possible effects of this optimization method on overall vaccination coverage, we applied the poisson participation probability regression to compare predicted participation in the vaccination campaign for the actual tent locations selected historically based on convenience to the optimally selected based on recursive replacement to maximize expected participation in the vaccination campaign based on the (Figure 5).

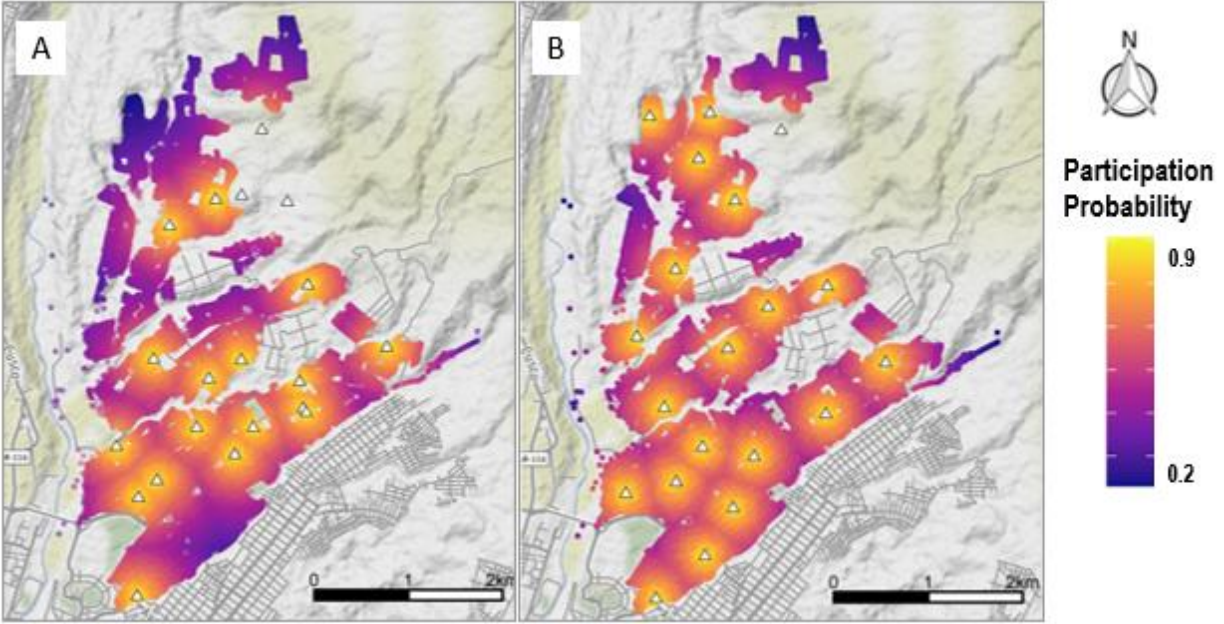

**Figure 5: Predicted vaccination campaign participation.** Panel A shows the data from the 2016 MDVC. Tent locations were selected based on convenience. The predicted coverage obtained from applying our participation probability regression was 43.18%. Panel B shows optimized placement of tents based on estimated probability of participation optimization (negative binomial regression). The predicted participation using this method was improved at 48.40%.

**DISCUSSION**

We present an application of the facility location problem that optimally places fixed location vaccination points for the annual mass dog rabies vaccination campaign in Arequipa, Peru. In line with previous studies examining barriers to vaccination (12,53–55), we found a significant negative association between walking distance from a vaccination location and household participation in the vaccination campaign. In order to maximize the coverage obtained from a fixed number of vaccination points in the city, we optimized the locations of the vaccination points under the framework that participation probability decreased with walking distance from the vaccination tent. We found the p-median method (which minimizes overall walking distance) did better than the p-center method and conventional methods to create the most even workload across vaccination points and increased overall estimated vaccination coverage. Compared to historical data with tents placed using conventional methods, we predict based on our Poisson regression that coverage would be increased by 11.2% if the p-median optimization algorithm of tents was applied.

This algorithm uses a data-driven approach to combat an ongoing rabies epidemic by optimally using limited resources to maximize vaccination coverage. The main effects we expect would be increased overall vaccination coverage and increased spatial evenness of coverage. The Pan-American Health Organization recommends that 80% of dogs get vaccinated annually to reach population immunity levels effective to eliminate canine rabies from a region (56). Our optimization algorithm helps get closer to this goal. The World Health Organization has set a goal of zero human deaths from rabies by 2030 (57). The most effective way to prevent canine-transmitted human rabies is to control rabies in the canine population through mass dog vaccination (58); however, this requires sustained yearly vaccination campaigns requiring the mobilization of health officials and dog owners (59). Fixed point vaccination locations in which dog owners bring their dogs to the vaccination point are much more cost effective than other methods such as door-to-door vaccination (60). Facility location optimization can help best locate the vaccination point locations; this is especially important for a neglected disease that suffers globally from underfunded control programs (61). Canine rabies control programs are underfunded globally and this is certainly the case in Arequipa, Peru. Control programs worldwide are under additional strains as the COVID-19 pandemic has diverted funds and resources away from other public health initiatives (15). The algorithm we present can help to best use limited resources. Furthermore, this application of the facility location problem can be used beyond rabies. Any mass vaccination campaign that sets up fixed location points to which people travel to for vaccinations could be enriched by this algorithm as long as distance is a barrier to access.

We developed a computational solution to optimally place vaccination points by minimizing overall walking distance. Based on our Poisson regression, we predict that this will increase overall vaccination coverage and vaccination coverage evenness in our study area. Our future directions include validating this method via a field trial. A limitation of our study is that we looked solely at walking distance from a vaccination point to determine the probability of MDVC participation. However, it is known that there are many factors that affect MDVC participation which we did not evaluate. Examining other factors contributing to MDVC participation would strengthen strategies developed to increase vaccination coverage. A second limitation of our algorithm is that it does not include the variation in workload created as a result of different household densities. A high workload could create lines at the vaccination sites increasing abandonment rates and generating a negative feedback in the system.

Rabies virus circulates at very low levels within the population (32). Studies have shown that sustained rabies virus transmission can persist in pockets of under-vaccinated dogs (35). This suggests that not only is a high vaccination coverage needed, but that this coverage is needed to be spread evenly over the population (34). By optimizing the location of vaccination points; access to the vaccine is balanced across the study area. When locations are selected based on convenience, neighborhoods that are more isolated and less convenient are disproportionately disenfranchised from access to rabies vaccinations. Using a data-driven approach to optimally place vaccination sites could increase the spatial homogeneity of vaccine coverage and increase the chances of control or elimination.

**ACKNOWLEDGEMENTS**

We gratefully acknowledge the members of the Zoonotic Disease Research Laboratory who helped with annual surveys of households in the study area, as well as the Ministry of Health who ran the annual mass dog vaccination campaigns and shared their expertise with our team.